\documentclass[a4paper]{article}
\usepackage{cite}
\usepackage{INTERSPEECH2019}

\title{An End-to-End Audio Classification System based on Raw Waveforms and Mix-Training Strategy}
\name{Jiaxu Chen*, Jing Hao*, Kai Chen, Di Xie, Shicai Yang, Shiliang Pu\thanks{*These first two authors contribute equally to this work.}}
\address{Hikvision Research Institute, Hangzhou, China}
\email{\{chenjiaxu, haojing, xiedi, yangshicai, pushiliang\}@hikvision.com}

\begin{document}

\maketitle
\begin{abstract}
Audio classification can distinguish different kinds of sounds, which is helpful for intelligent applications in daily life. However, it remains a challenging task since the sound events in an audio clip is probably multiple, even overlapping. This paper introduces an end-to-end audio classification system based on raw waveforms and mix-training strategy. Compared to human-designed features which have been widely used in existing research, raw waveforms contain more complete information and are more appropriate for multi-label classification. Taking raw waveforms as input, our network consists of two variants of ResNet structure which can learn a discriminative representation. To explore the information in intermediate layers, a multi-level prediction with attention structure is applied in our model. Furthermore, we design a mix-training strategy to break the performance limitation caused by the amount of training data. Experiments show that the mean average precision of the proposed audio classification system on Audio Set dataset is 37.2\%. Without using extra training data, our system exceeds the state-of-the-art multi-level attention model.
\end{abstract}

\section{Introduction}

Sound is an indispensable medium for information transmission of the surrounding environment. When some sounds happen, such as baby crying, glass breaking and so on, we usually expect that we can ``hear'' the sounds immediately even if we are not around. In this case, audio classification which aims to predict whether an acoustic event appears has gained great attention in recent years. It has many practical applications in remote surveillance, home automation, and public security.

In real life, an audio clip usually contains multiple overlapping sounds, and the types of sounds are various, ranging from natural soundscapes to human activities. It is challenging to predict the presence or absence of audio events in the audio clip. Audio Set \cite{7} is a common large-scale dataset in this task, which contains about two million multi-label audio clips covering 527 classes. Recently, some methods have been proposed to learn audio tags on this dataset. Among them, a multi-level attention model \cite{3} achieved the state-of-the-art performance based on \cite{2}, which outperforms Google's baseline \cite{1}. However, the shortcoming of these models is that the input signal is the published bottleneck feature \cite{1}, which causes information loss. Considering that the actual length of sound events is different and the handcrafted features may throw away relevant information at a short time scale, raw waveforms containing richer information is a better choice for multi-label classification. In audio tagging task of DCASE 2017, 2018 challenge \cite{4}, some works \cite{5,6} combined handcrafted features with raw waveforms as input signals on a small dataset consisting of 17 or 41 classes. To our knowledge, none of the works proposes an end-to-end network taking raw waveforms as input in Audio Set classification task.

In this paper, we propose a classification system based on two variants of ResNet \cite{13} which directly extracts features from raw waveforms. Firstly, we use a one-dimension (1D) ResNet for feature extraction. Then, a two-dimension (2D) ResNet with multi-level prediction and attention structure is used for classification. For obtaining better classification performance further, a mix-training strategy is implemented in our system. In this training process, the network is trained with mixed data which extends the training distribution, and then transferred to target domain using raw data.

In this work, the main contributions are as follows:
\begin{enumerate}
\item A novel end-to-end Audio Set classification system is proposed. To the best of our knowledge, it is the first time to take raw waveforms as input on Audio Set and combine 1D ResNet with 2D ResNet for feature extraction at different time scales.
\item A mix-training strategy is introduced to effectively improve the utilization of limited training data. Experiments show that it is powerful in multi-label audio classification compared to the existing data augmentation methods.
\end{enumerate}

The remainder of this paper is organized as follows. Section 2 presents related works. Section 3 introduces the proposed network architecture and mix-training strategy. Section 4 shows the experimental results and analysis. Finally, section 5 gives the conclusion of this paper.

\begin{figure*}[t]
  \centering
  \includegraphics[width=\linewidth]{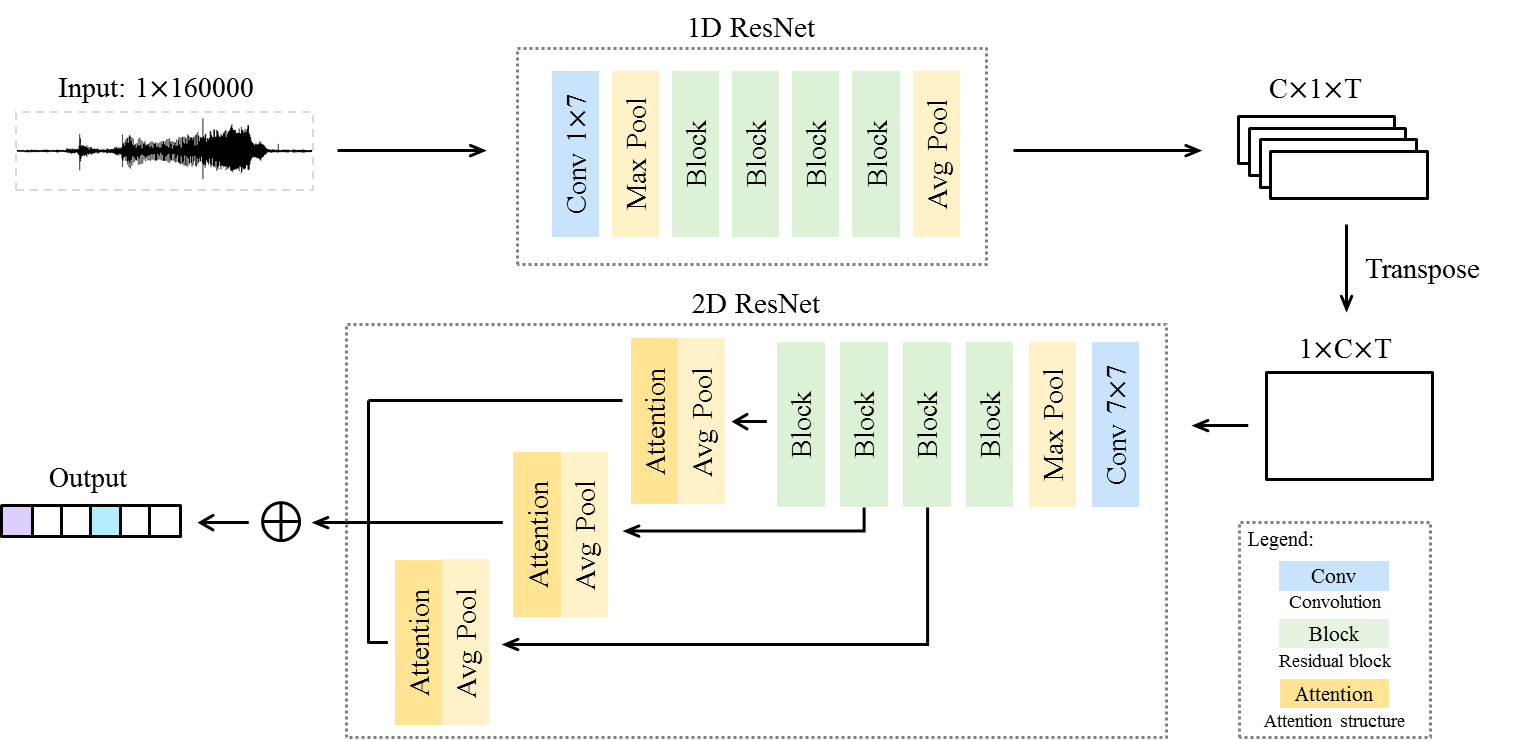}
  \caption{Architecture of the end-to-end audio classification network. The raw waveform (1D vector) is the input signal. First, a 1D ResNet is applied to extract audio features. Then, the features are transposed from $C \times 1 \times T$ to $1 \times C \times T$. Finally, a 2D ResNet with multi-level prediction structure performs audio classification. The output of the network has multiple labels and is the mean of the multi-level prediction results. The Block is composed of $n$ bottleneck blocks, where $n$ is related to the number of layers in ResNet.}
  \label{fig:network}
\end{figure*}

\section{Related works}

Deep learning has achieved success in audio classification tasks. The majority of these works \cite{Takahashi2017, 1, Karol2015, Chen2018} classify the features designed by humans, such as log-mel feature and MFCC. Recently, some works learning the audio tags directly from the raw waveforms present high performance. \cite{DaiW2016} proposed a time-domain network based on the 1D raw waveform. The results in \cite{12, ZhuB2018} demonstrated that the feature directly learned from the raw waveform is sufficient for audio classification problems. However, these works are designed specially for single-label audio classification. For multi-label classification, we input raw waveforms to a multi-level prediction with attention structure which can perceive the sound at different time scales.  

The lack of data is a crucial issue in audio classification. Meanwhile, tagging the sound events consumes amounts of manpower. To address this problem, two kinds of approaches have been proposed: one efficiently making use of limited data \cite{Takahashi2016, 14, ZhangH2018, 10, 11} and the other based on additional data \cite{Yusuf2016, Kumar2017, 2}. In the first type of method, Takahashi et al. \cite{Takahashi2016} mixed two different sounds belonging to one class to extend the training distribution. Mixup \cite{ZhangH2018} interpolated the training data and trained the model to output the mixing ratio. However, these augmentation methods are not suitable for a multi-label dataset, such as Audio Set. In the second type, Kumar et al. \cite{Kumar2017} pre-trained the model by an additional large audio dataset. Kong et al. \cite{2} applied the rich sound representation learned on YouTube-100M \cite{1} to classify Audio Set.

Different from all the approaches above, we propose a simple training strategy to alleviate the need of training data in audio classification tasks. Specifically, we train the network using both mixed data and raw data, and only focus on the event categories other than the mixing ratio in audio clips. The proposed method can be applied in both single-label and multi-label audio classification. What's more, there is no requirement of pre-training stage with an additional large dataset. The experiments in Section 4 demonstrate the superiority of the proposed training strategy. Our network trained with only Audio Set presents a comparable performance to the state-of-the-art method \cite{3} which uses the extra knowledge of YouTube-100M.

\begin{figure*}[t]
  \centering
  \includegraphics[width=\linewidth]{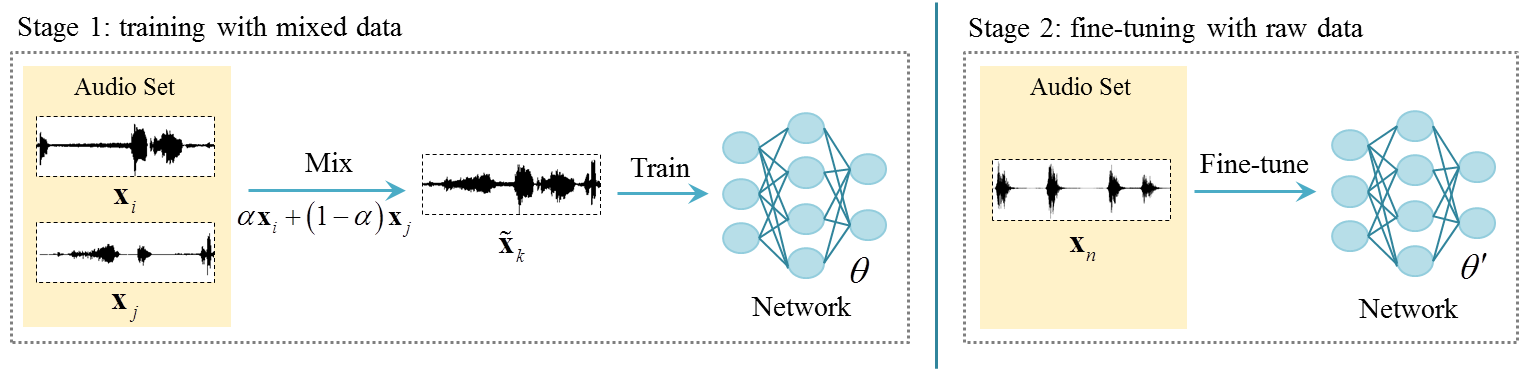}
  \caption{Mix-training strategy. We first train the network $\theta$ on mixed data ${\bf{\tilde x}}_k$ and then fine-tune it to target domain $\theta'$ using raw data ${\bf{x}}_n$. ${\bf{x}}_i$, ${\bf{x}}_j$ and ${\bf{x}}_n$ are raw data randomly selected from Audio Set and ${\bf{\tilde x}}_k$ is the linear interpolation of audio clip pairs $\left( {{\bf{x}}_i ,{\bf{x}}_j } \right)$ at a random ratio $\alpha \in \left( {0,1} \right)$.}
  \label{fig:training}
\end{figure*}

\section{Raw-waveforms-based network and mix-training strategy}

In this section, we describe the raw-waveforms-based end-to-end network and mix-training strategy. Compared to handcrafted features, raw waveforms preserve more complete information of audio clips. Thus, we propose a novel network based on two variants of ResNet structure, where feature extraction and classification are jointly performed on raw waveforms. Furthermore, training the end-to-end network with a novel mix-training method yields better performance.

\subsection{Raw-waveforms-based network}

Figure \ref{fig:network} illustrates the detailed architecture of our network. The network is split into two stages: the top stage computes frequency-like features and the bottom stage classifies the extracted features.

The raw waveform, which can be represented as 1D vector, is the input signal of our network. At the top stage, a 1D time-domain residual network, called 1D ResNet, is applied to extract audio features. At the bottom stage, a 2D residual network with multi-level prediction structure, called 2D ResNet, performs audio classification. As shown in Figure \ref{fig:network}, the shape of the output feature in the top stage is defined as 
$C \times 1 \times T$, where $C$ is the number of channels and $T$ is the time dimension. Specifically, the output feature is frequency-like and consists of $C$ features with a size of $1 \times T$. In order to convolve in both time and frequency, we transpose the feature from $C \times 1 \times T$ to $1 \times C \times T$ and then input the 2D frequency-like map to 2D ResNet. In the bottom stage, a multi-level prediction structure is applied to make use of the information in intermediate layers, which can also prevent gradient vanishing.  In this way, we can obtain three prediction results after the 2th, 3th, 4th Block respectively. In Figure \ref{fig:network}, the basic unit of the Block is building block or bottleneck block, where the type and the number of the units are determined by the number of layers in ResNet. The overall optimization objective of the network is to minimize the mean of multi-level prediction loss. Moreover, to quantify the importance of audio segments containing audio events, we adapt the attention structure \cite{3} in the multi-level prediction. In our work, the attention structure consists of two attention modules, where one is applied after Block and the other is applied after the two fully connected (FC) layers following the Block. Then, the two predictions of the attention modules are concatenated to a vector as the output of the attention structure.

\subsection{Mix-training strategy}

As is known, the performance of a network greatly depends on the amount and quality of training data. A mix-training strategy (see Figure \ref{fig:training}) is introduced to improve the utilization of the limited training set. It can be concluded as a two-step training process: first, the model parameter is optimized with the mixed audio clips; then, we use the raw audio data to fine-tune the model parameter from $\theta$ to $\theta'$.

The sound collected from the real environment, such as the sample in Audio Set, may contain more than one audio event. Humans can distinguish such sounds. Therefore, mixing two different audio clips to produce one new audio clip is a reasonable data augmentation method. In the first training step, we train the proposed network on the mixed data. Let  $\left( {{\bf{x}}_i ,{\bf{y}}_i } \right)$ and $\left( {{\bf{x}}_j ,{\bf{y}}_j } \right)$ be a pair of randomly selected training samples in the form of $\left( {{raw\ waveform} ,{label} } \right)$, and $\left( {{\bf{\tilde x}}_k ,{\bf{\tilde y}}_k } \right)$ be mixed data. ${\bf{y}}_i$ and ${\bf{y}}_j$ are both multi-hot vectors composed of 0 and 1. A multi-label audio clip can be seen as the mixture of multiple single-label audio clips. Due to the nature of multi-label audio, it's difficult to predict the mixing ratio when mixing two multi-label audio clips. Thus, mixup \cite{ZhangH2018} which predicts ratio label  ${\bf{\tilde y}}_k  = \alpha {\bf{y}}_i  + \left( {1 - \alpha } \right){\bf{y}}_j$ is not suitable for multi-label classification. Different from mixup, we train the network to output only the event categories of mixed data. In our method, ${\bf{x}}_i$ and ${\bf{x}}_j$ can contain multiple labels. The mixed training data $\left( {{\bf{\tilde x}}_k ,{\bf{\tilde y}}_k } \right)$ is generated as:
\begin{equation}
{\bf{\tilde x}}_k  = \alpha {\bf{x}}_i  + \left( {1 - \alpha } \right){\bf{x}}_j 
\end{equation}
\begin{equation}
{\bf{\tilde y}}_k  = sign\left( {{\bf{y}}_i  + {\bf{y}}_j } \right)
\end{equation}
where the mixed audio ${\bf{\tilde x}}_k$ is the linear interpolation of ${\bf{x}}_i$ and ${\bf{x}}_j$ at a random ratio $\alpha  \in \left( {0,1} \right)$, and the corresponding label ${\bf{\tilde y}}_k$ contains all the labels in  ${\bf{x}}_i$ and ${\bf{x}}_j$. $sign\left(  \cdot  \right)$ is the elementwise operation that extracts the sign of each element of a vector.

Classifying the mixed audio can also be seen as a multi-label classification problem. Thus, we optimize the model parameter $\theta$ using sigmoid cross entropy loss:
\begin{equation}
L =  - \frac{1}{K}\sum\limits_{k,n} {\left[ {\left( {1 - \tilde y_{kn} } \right)\log \left( {1 - t_{kn} } \right)} \right. + \left. {\tilde y_{kn} \log t_{kn} } \right]} 
\end{equation}
\begin{equation}
{\bf{t}}_k  = f_\theta  \left( {{\bf{\tilde x}}_k } \right) = \left[ {t_{k1} ,t_{k2} , \cdots ,t_{kN} } \right]
\end{equation}
where ${\bf{t}}_k$ is the prediction result of  ${\bf{\tilde x}}_k$, $t_{kn}$ and $\tilde y_{kn}$ are the predicted probability and real probability of event $n$ happened in ${\bf{\tilde x}}_k$. $K$ and $N$ are the number of audio files and classes.

Mixing operation increases the variety of training data, and thereby leads to better performance. However, the distribution of the mixed data produced by linear interpolation is different from that of the raw data. Owing to domain shift \cite{Torralba2011}, the model $\theta$ trained on the mixed data could not generalize perfectly to the original classification task. The mixed data blending more events than raw data still keeps the characteristic of audio. Therefore, the representation learned from the mixed data can be transferred to the raw data. Through fine-tuning the trained model $\theta$ with the raw training data, we can transfer the network to the original audio classification task.

\section{Experiments}

\begin{table*}[t]
  \caption{Comparisons of results of different works on Audio Set classification.}
  \label{tab:all}
  \centering
  \begin{tabular}{llllll}
    \toprule
    \textbf{Method} & \textbf{mAP} & \textbf{AUC} & {\textbf{d-prime}} & \textbf{Input} & \textbf{Pre-train (Dataset)} \\
    \midrule
    {\cite{1} Benchmark} & {0.314} & {0.959} & {2.452} & {Log-mel spectrogram} & {Yes (YouTube-100M)} \\   
    {\cite{2} Single-level attention} & {0.327} & {0.965} & {2.558} & {Bottleneck features} & {Yes (YouTube-100M)} \\  
    {\cite{3} Multi-level attention} & {0.360} & {\bf 0.970} & {\bf 2.660} & {Bottleneck features} & {Yes (YouTube-100M)} \\  
    {Ours w/o mix-training} & 0.351 & 0.962 & 2.510 & {Raw waveforms} & {No} \\
    {Ours} &{\bf 0.372} & {0.968} & {2.614} & {Raw waveforms} & {No} \\
    \bottomrule
  \end{tabular}
\end{table*}

\subsection{Dataset}

Audio Set \cite{7} is a human-labeled dataset for audio events and contains over 2 million 10-sec audio clips excerpted from YouTube videos covering 527 classes. Each audio clip in this dataset may contain multiple labels. Audio Set can be divided into a balanced training set about 20,000 segments, an unbalanced training set with different number of audio clips per class about 2 million segments and a balanced evaluation set about 20,000 segments. For Audio Set classification, the previous works \cite{2,3} take the published bottleneck features as input and the dimension of each feature is 10 $\times$ 128. Specifically, the features are extracted from a ResNet-50 model pre-trained on log-mel spectrograms of 70 million samples in YouTube-100M \cite{1}. Unlike these works, we use only Audio Set other than much larger YouTube dataset to train the classification model. In our network, the input is the raw waveform with a dimension of 1 $\times$ (10 $\times$ 16000),  which indicates the audio clip length is 10 seconds and the sampling frequency is 16 kHz.

\subsection{Details}

As shown in Figure \ref{fig:network}, we propose a two-stage ResNet architecture for Audio Set classification taking raw waveforms as input. In this architecture, 1D ResNet-18 extracts discriminative frequency-like features and 2D ResNet-50 combining multi-level prediction is used to classify the extracted features. The multi-level prediction is implemented by applying attention structure after the 2th, 3th, 4th bottleneck block, respectively. In the attention structure, the hidden units of the fully connected layer are 600 and the activation function is ReLU. In our experiments, we apply the mix-training strategy to train the two-stage ResNet architecture. We use Adam optimizer \cite{15}, and the learning rate is $3 \times 10^{-4}$ and  $3 \times 10^{-5}$ in training and fine-tuning step. In the mix-training process, we experimentally find that the mixed ratio $\alpha  \in \left( {0.4,0.6} \right)$ can achieve better performance. To solve the highly unbalanced problem of Audio Set, we use a data balancing strategy where the training samples of each class appear with the same frequency in a batch. 

Like \cite{1,2,3}, we evaluate our model with Google's benchmark metrics: mean average precision (mAP), area under curve (AUC) and d-prime. The higher the values of these metrics, the better the audio classification performance.

\subsection{Analysis}

\subsubsection{Comparisons of training strategies}
In this subsection, we employ the balanced training set in Audio Set to explore the effectiveness of the mix-training strategy.
The results evaluated on the balanced evaluation set are summarized in Table \ref{tab:training_bal}. When mixup \cite{ZhangH2018} is applied to train the model, the performance is improved. However, the ratio label ${\bf{\tilde y}}_k  = \alpha {\bf{y}}_i  + \left( {1 - \alpha } \right){\bf{y}}_j$ used in mixup limits the improvement in multi-label classification. The mixed label ${\bf{\tilde y}}_k  = sign\left( {{\bf{y}}_i  + {\bf{y}}_j } \right)$ generated by our mix-training strategy presents better performance. Fine-tuning, adapting the network to the original data distribution, leads to further performance improvement. The mix-training strategy is a better choice for multi-label audio classification.

\begin{table}[!htbp]
  \caption{Comparisons of training strategies on the balanced training set containing about 20,000 segments. ``No'' means that we train the network without any training strategy.}
  \label{tab:training_bal}
  \centering
  \begin{tabular}{llll}
    \toprule
    \textbf{Training strategy} &\textbf{mAP} & \textbf{AUC} & \textbf{d-prime} \\
    \midrule
    No & 0.231 & 0.94 & 2.22 \\
    \cite{ZhangH2018} mixup & 0.270 & 0.94 & 2.25  \\
    mix-training w/o fine-tuning & 0.276 & 0.94 & 2.27 \\
    mix-training & 0.296 & 0.95 & 2.38 \\
    \bottomrule
  \end{tabular}
\end{table}

\subsubsection{Comparisons of parameters}

Table \ref{tab:size} shows the comparisons of results with different output sizes of 1D-ResNet and batch sizes. These experiments are based on the unbalanced training set containing about 2 million segments. As described in Section 3.2, the output size of 1D-ResNet is  $C \times 1 \times T$, where $C$ is the number of features with a size of $1 \times T$. In our experiments, we observe that the larger the size of frequency-like features, the better the audio classification performance. The reason is that the larger features can retain richer information. Experimentally, the larger batch size which can reduce the variance of stochastic gradients is also helpful to improve the performance. Limited by GPU memory, the output size of 1D-ResNet is $128 \times 1 \times 2500$ and the batch size is $256$ in our experiments.

\begin{table}[!htbp]
  \caption{Comparisons of results with different output sizes of 1D-ResNet and batch sizes.}
  \label{tab:size}
  \centering
  \begin{tabular}{lclll}
    \toprule
    \textbf{Output size} & \textbf{Batch size}& \textbf{mAP} & \textbf{AUC} & {\textbf{d-prime}} \\
    \midrule
    {$128 \times 1 \times 1250$} & {192} & {0.349} & {0.961} & {2.500} \\   
    {$128 \times 1 \times 2500$} & {192} & {0.372} & {0.965} & {2.558} \\  
    {$128 \times 1 \times 2500$} & {256} & {0.372} & {0.968} & {2.614} \\  
    \bottomrule
  \end{tabular}
\end{table}

\subsubsection{Results}

In Table \ref{tab:all}, we compare the classification results of different works on Audio Set. The mAP of the state-of-the-art classification model \cite{3} on Audio Set is 0.360, which outperforms Kong et al. \cite{2} of 0.327 and Google's baseline \cite{1} of 0.314. To our knowledge, all the three works pre-trained their models with the YouTube-100M much larger than Audio Set and none of them is an end-to-end system. In our experiments, we use only Audio Set to train the proposed end-to-end model (see Figure 1) by taking raw waveforms as input without any pre-training process. Furthermore, the mix-training strategy is adopted to improve the performance. Without the assistance of extra data but Audio Set, our method achieves a 1.2\% improvement in mAP over the state-of-the-art method. 

Specifically, the last two lines in Table \ref{tab:all} shows the classification results of our proposed network with and without mix-training strategy. As shown in Table  \ref{tab:all}, the values of mAP, AUC and d-prime are all higher if we use mix-training strategy than not, which proves that the mix-training strategy is helpful to improve the classification performance. Compared to Table \ref{tab:training_bal}, the degree of improvement is smaller in Table  \ref{tab:all}. That's because the good performance depends on a sufficient amount of training data and the smaller training set can get more performance gain from the data augmentation.

\section{Conclusion}

In this paper, we propose a raw-waveforms-based end-to-end audio classification system with a mix-training strategy. Inputting raw waveforms, the system performs both feature extraction by 1D ResNet and classification by 2D ResNet. It is the first attempt to take raw waveforms as input in Audio Set classification, which retains rich information and makes end-to-end possible. Furthermore, we apply a mix-training strategy to train the model with both mixed data and raw data. The strategy is simple and powerful to improve the performance on the limited training data. Experiments show that the proposed system trained with only Audio Set outperforms the state-of-the-art multi-level attention model using the knowledge pre-trained on YouTube-100M. In future work, the mix-training strategy could provide an approach to alleviate the data limitation in other tasks, such as speech recognition and image classification.

\bibliographystyle{IEEEtran}

\bibliography{mybib_new_2}


\end{document}